\documentclass{article}

\usepackage{PRIMEarxiv}

\usepackage[utf8]{inputenc} 
\usepackage[T1]{fontenc}    
\usepackage{hyperref}       
\usepackage{url}            
\usepackage{booktabs}       
\usepackage{amsfonts}       
\usepackage{nicefrac}       
\usepackage{microtype}      
\usepackage{lipsum}
\usepackage{fancyhdr}       
\usepackage{graphicx}       
\graphicspath{{media/}}     

\title{Fast CT Anatomic Localization Algorithm
}

\author{
  Amit Oved \\
  \texttt{amitoved@gmail.com} \\
}

\begin{document}
\maketitle

\begin{abstract}
Automatically determining the position of every slice in a CT scan is a basic yet powerful capability allowing fast retrieval of region of interest for visual inspection and automated analysis. Unlike conventional localization approaches which work at the slice level, we directly localize only a fraction of the slices and and then fit a linear model which maps slice index to its estimated axial anatomical position based on those slices. The model is then used to assign axial position to every slices of the scan.
This approach proves to be both computationally efficient, with a typical processing time of less than a second per scan (regardless of its size), accurate, with a typical median localization error of 1 cm, and robust to different noise sources, imaging protocols, metal induced artifacts, anatomical deformations etc.
Another key element of our approach is the introduction of a mapping confidence score. This score acts as a fail safe mechanism which allows a rejection of unreliable localization results in rare cases of anomalous scans. 
Our algorithm sets new State Of The Art results in terms of localization accuracy. It also offers a decrease of two orders of magnitude in processing time with respect to all published processing times. It was designed to be invariant to various scan resolutions, scan protocols, patient orientations, strong artifacts and various deformations and abnormalities. Additionally, our algorithm is the first one to the best of our knowledge which supports the entire body from head to feet and is not confined to specific anatomical region. This algorithm was tested on thousands of scans and proves to be very reliable and useful as a preprocessing stage for many applications. 

\end{abstract}

\section{Introduction}
Computer Aided Diagnosis (CAD) systems have rapidly entered the radiologist work-flow and are estimated to have an increasing role in the upcoming years as more and more companies develop tools for measurements, detection and classification of anatomical structures, abnormality and pathology in different modalities. One of the widely used imaging technology is the Computed Tomography (CT) scan \cite{health_at_glance-2017}. Unlike 2D imaging modalities, CT scan data is volumetric and is typically composed of hundreds or even thousands of 2D image slices. As a result, the file size of a typical scan is in the order of hundreds of megabytes and therefore takes considerable amount of time and resources to process. This fact, combined with the fact that future Picture Archiving and Communication Systems (PACS) and radiologist workstations are expected to apply many different algorithms on a single scan is expected to lead to substantial processing time per scan.

Most CT based algorithms and radiologists spend considerable resources on first locating the organ or region of interest before analyzing it. Considering, for example a liver lesion detection algorithm operating on a chest-abdomen CT scan. Such an algorithm first needs to process the entire scan in order to localise the liver. Once the liver is located, the lesions detection algorithm can then process the sub volume containing the liver which is much smaller than the original volume and look for suspicious lesions. Such an algorithm could be much faster if it was provided with the sub-volume containing the axial region of interest where the liver is located instead of the entire scan. While DICOM meta-data provides information of the scan’s anatomical region, it is error prone and does not provide the necessary level of granularity \cite{DICOM-Quality} required by organ specific algorithms. 
In order for such an approach to actually reduce the overall computational load, we suggest a fast preprocessing stage which assigns every slice of the scan a unique number which represents its position along the axial dimension of the human body. Once the entire volume is indexed, different sub-volumes can be delivered to different algorithms based on their anatomical region of interest. In this paper we present an algorithm which maps every slice in a CT scan to it's normalized axial position which to the best of our knowledge is the most accurate and the fastest. Also, it was designed to support a variety of scan resolutions, scan protocols, patient orientations, strong artifacts and various deformations and abnormalities. Additionally, our proposed algorithm is the first one to the best of our knowledge which supports the entire body from head to feet.

\section{Previous work}
Existing approaches on automatic positioning of CT scans can be divided into two main groups: hand crafted feature based and learned feature based. 
In a hand crafted feature based approach, a feature vector is calculated for each image based on some heuristics. For instance, the input image is divided into a set of smaller segments by a 2D grid where the grid can be Cartesian \cite{CT_slice_localization}, \cite{Comparing_axial_CT} or polar \cite{Radial_Image_Descriptors}. For every segment, a local feature vector which is usually the distribution of gray levels is calculated. The local feature vectors are concatenated to form a single global feature vector which is then mapped to a scalar indicating the estimated position of that slice along the axial dimension of the body. This mapping function can be linear or a more sophisticated one \cite{deep_learning_based}. While most studies treat each slice separately, the authors in \cite{Position_prediction} aggregated information from neighboring slices by weighted averaging the global feature vectors of several slices in order to increase accuracy and robustness to noise. 
In the learned feature based approach which is considered better at generalizing and dealing with more complex and noisy inputs, both image representation and the mapping function which maps the feature vector to a position along the axial dimension of the body are learned. 
In \cite{ANATOMY-SPECIFIC}, the authors trained a CNN to classify an input slice region to one out of five possible regions. \cite{UNSUPERVISED} trained a convolution based regression network to predict a continuous value which indicates the position of a slice along the axial dimension of the human body ranging from the chest to the pelvis. 
Our approach belongs to the second group of approaches and is also based on a CNN to predict the position of the input image anywhere along the axial dimension of the human body. 
The main novelty in our approach is the aggregation of the CNN based localization with the known position of every slice in a real world coordinate system. This fusion of information results in a very accurate and fast positioning of an entire scan.

\section{Methods}
\subsection{Normalized coordinate system}
In order to establish a normalized 1-dimensional coordinate system \cite{CT_slice_localization}, \cite{Estimating_the_body_portion} which is invariant to the patient’s height and scan resolution, we chose to scale the entire human body to a continuous range of values ranging from 0 to 99 such that the scaled position of the tip of the head and the sole of the foot are set to 0 and 99, respectively. Additionally, we assign to nine other distinct anatomical landmarks nine values based on their relative position in the human body. The different anatomical positions and their corresponding values appear in Table~\ref{tab1}. Once this coordinate system is established, the problem of localizing every slice in a CT scan is reduced to the problem of mapping every slice to a scalar in the range [0, 99].
The labeling of our data-set was performed the following way: given a CT scan, a trained annotator located the slices showing the superior (uppermost) and inferior (lowermost) visible anatomical landmarks out of the eleven landmarks defined in Table 1. Those two slices were assigned with the landmarks’ corresponding values. The remaining slices between in between were assigned with the linear interpolation of the values assigned to the two extreme slices. This labeling method is very efficient since the identification of two landmarks in a given scan, leads to an automatic labeling of all slices between the two.

\begin{table}
\centering
\begin{tabular}{|c|c|}
\hline
Anatomical landmark &  Relative location\\
\hline
Superior aspect of th skull & 0.0\\
Inferior cerebellum & 10.9\\
Hyoid bone & 12.6\\
Superior sternum & 18.9\\
Carina & 21.1\\
Inferior aspect of the heart & 28.0\\
Lower portion of the 12th rib & 36.6\\
Superior ilium & 40.0\\
Lesser trochanter & 51.4\\
Patellas & 71.4\\
Sole of the foot & 100.0\\
\hline
\end{tabular}
\vspace{5mm} 
\caption{Anatomical landmarks and their positions. Eleven landmarks selected which are distinct and are distributed along the entire human body.}\label{tab1}
\end{table}

\subsection{Slice level localization} \label{classifier}
\begin{figure}
\begin{center}
\includegraphics[width=1.0\textwidth]{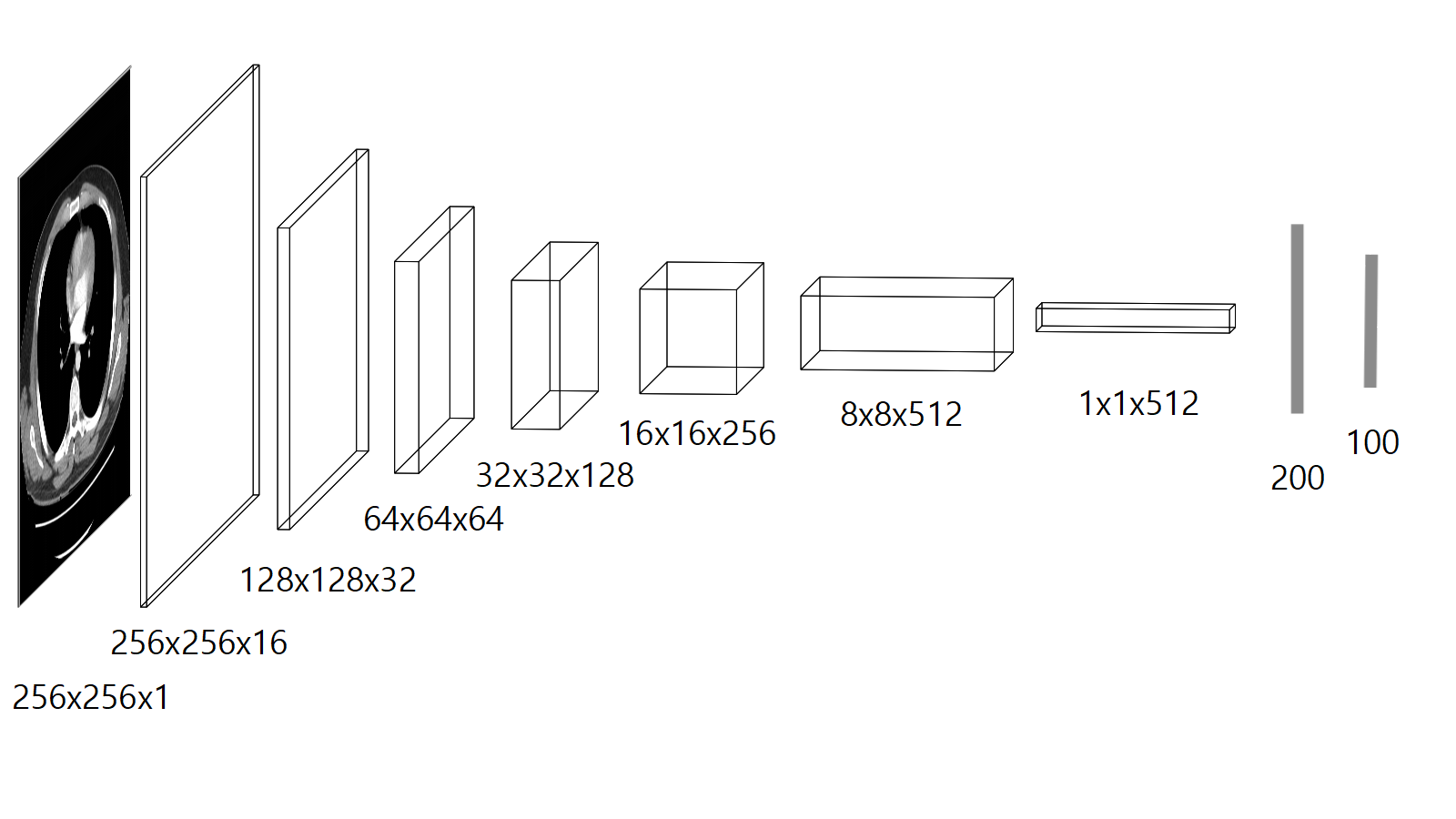}
\caption{Classification architecture. The classifier utilized for determining the position (class) of an input slice} 
\label{architecture}
\end{center}
\end{figure}

For the slice localization (classification), we used a simple CNN (Fig.~\ref{architecture}) which is composed of a sequence of 3x3 convolution layers followed by 2x2 max pooling layers ending with two fully connected layers. The network's input is a single channel 256x256 image and its output is a softmax-normalized vector of length 100. Preprocessing included only the resizing of the original 512x512 CT slices to 256x256 slices using bi-linear interpolation. The original pixel values which indicate Hounsfield units were not rescaled. The labeling of our data-set was performed the following way: given a CT scan, a trained annotator located the slices showing the superior (uppermost) and inferior (lowermost) visible anatomical landmarks out of the 11 landmarks defined in Table~\ref{tab1}. Those two slices were assigned with the landmarks' corresponding values. The remaining slices between the two extremes were assigned with the linear interpolation of the extreme slices' values. This labeling method is very efficient since the identification of two extreme landmarks in a given scan, leads to an automatic labeling of all slices between the two.

\subsection{Training} 
The training set included 248 scans of 124 adult males and 124 adult females. The scans were chosen with an emphasis to include a variety of ages, scan protocols, anatomical deformations, abnormalities and artifacts inducing objects (pacemakers, screws, metallic hip and knee joints, amalgam fillings etc.). Those scans produced a collection of roughly 58,000 labeled slices as explained in \ref{classifier}, which were sub-sampled so that only the first out of every three consecutive slices was included in the training set resulting in 19287 slices. The validation set was curated by the same criteria and included 26 scans of 13 adult males and 13 adult females. After the same sub-sampling process, it included 2355 slices. 
Training was performed for 50 epochs using  Adam optimizer and categorical cross-entropy loss. The learning rate was $10^{-4}$ for the first 30 epochs, $10^{-5}$ for the next 10 epochs and $10^{-6}$ for the last 10 epochs. Data augmentation included zoom, rotation, shear, horizontal and vertical flips.

\subsection{Scan level localization} \label{Entire_scan}
Typical CT scan is essentially a collection of 2D images ordered by their relative position is space. This property can be exploited to dramatically reduce the amount of computation required to localize an entire scan and to substantially increase to localization accuracy and robustness to different noise sources compared to conventional approaches which treat every slice separately. Since all axial slices in the human body are mapped to the [0, 99] interval in a linear fashion, a mapping of an entire axial CT scan can be approximated by a linear model which maps slice index to position. It is worth noting that by modeling this mapping by a linear model, we assume humans have roughly the same proportions in the axial dimension. This assumption seems to be a valid assumption for the required accuracy levels in a typical use case and is supported by our accuracy tests. Ideally, this means that direct positioning of just two slices per scan is sufficient to map the entire scan regardless of it's size. In realistic scenarios, more slices are required to reliably estimate the linear mapping model in a way that is robust to small positioning inaccuracies and occasional strong outliers. Based on experiments with various scan regions and protocols we set the number of slices used to estimate the linear model to be 30, regardless of the number of slices in the volume. These slices are uniformly sampled from the entire scan. In order to estimate the model's parameters, a RANSAC algorithm \cite{RANSAC} which is considered to be robust to noise and outliers is used. This method not only reduces the amount of processing required localize an entire scan but also acts as a strong denoising mechanism. Additionally, the linearity assumption is also used to identify anomalies. Namely, low fitting score of the RANSAC algorithm indicates a possible problem with the localization process or the scan itself.

\section{Results} 
\subsection{Slice level performance} \label{classifier_performance}
\begin{figure}[h]
\begin{center}
\includegraphics[width=0.5\textwidth]{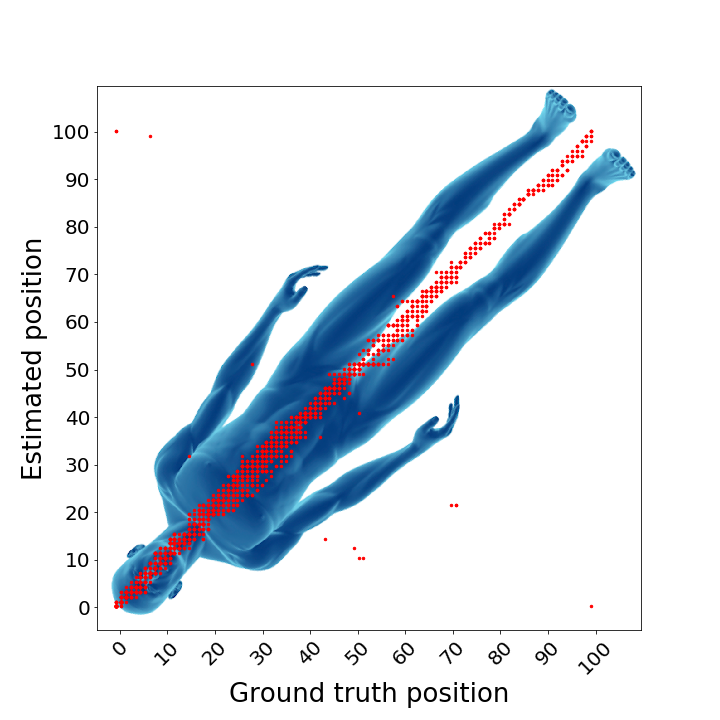}
\caption{Predicted label vs. ground truth labels of the entire validation set. Deviations of the diagonal indicate positioning errors. Those relatively large errors and the few outliers are later reduced as explained in \ref{Entire_scan}} 
\label{human}
\end{center}
\end{figure}

\begin{figure}[h]
\begin{center}
\includegraphics[width=0.8\textwidth]{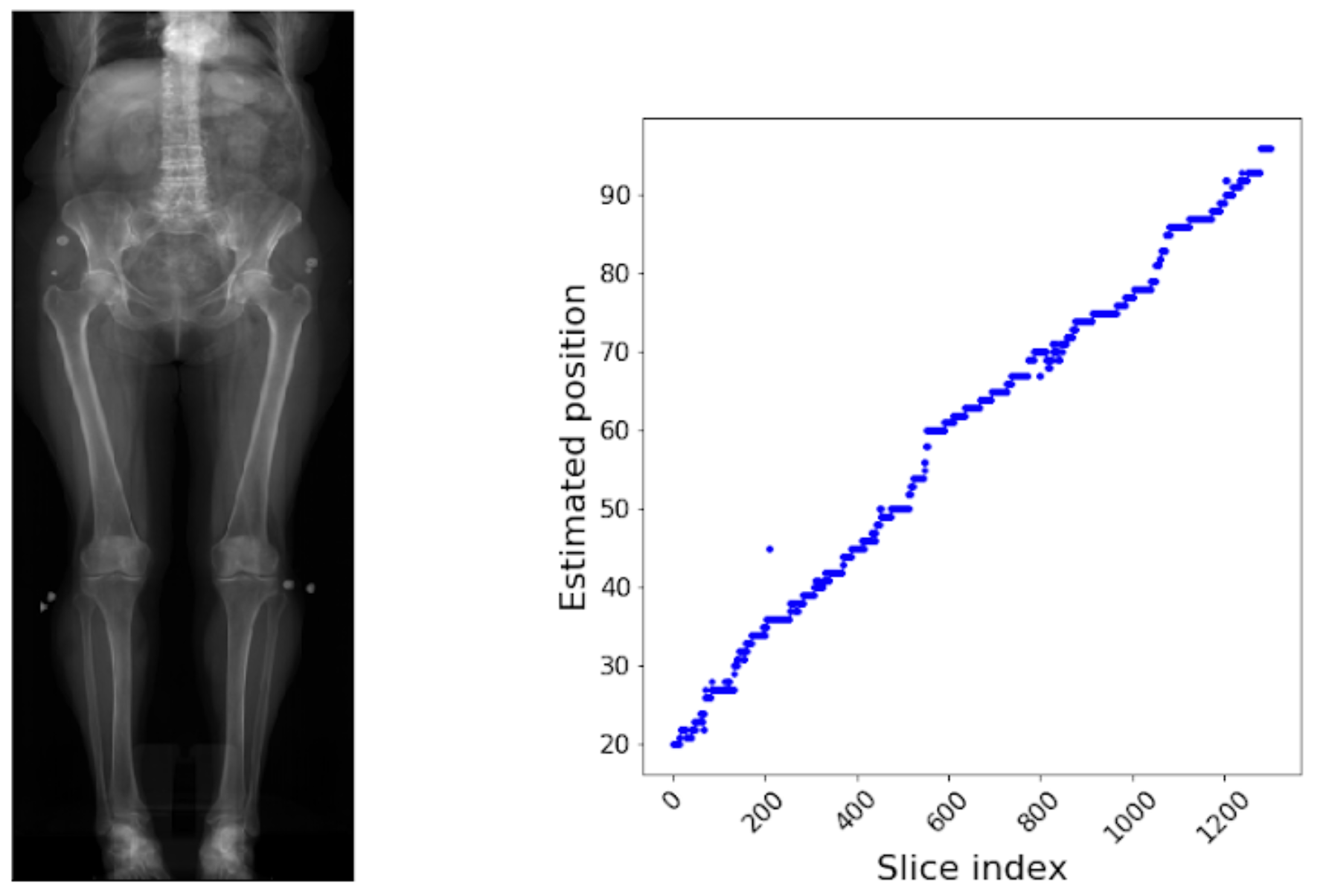}
\caption{Localization results of ~1300 slices} 
\label{loc_1300}
\end{center}
\end{figure}

 Fig.~\ref{human} determines the predicted class labels versus the ground truth labels for the entire validation set. Error analysis shows median and average errors of 1.0 and 1.4 units, respectively. Since the entire human body is mapped to a hundred possible locations and since the average adult height is roughly 170 cm, an error of one unit is equivalent to roughly 1.7 cm. This means that the median and average errors are equivalent to 1.7 and 2.38 cm respectively. It is important to bare in mind that the above values describe the classifier error rates, or the localization errors of a single slice without the context of the entire scan. For an entire scan localization (explained below), the error rates decrease. Fig.~\ref{loc_1300} shows a scan of ~1300 slices and their estimated locations. This result shows that a simple linear model which is based on a handful of slices can replace the costly processing of 1300 slices at a fraction of the time.

\subsection{Scan level performance} 
We performed two tests to analyze the accuracy of the algorithm. In test 1, a set of 104 scans of different sex, anatomical regions, protocols and resolutions was visually inspected in order to generate a set of {landmark, slice number} pairs per scan out of 12 possible landmarks. The scans were then processed by the suggested approach and the normalized position for every {landmark, slice number} pair was kept. The values were then grouped by landmarks and for each landmark, the average value, mean absolute error and median absolute errors were calculated. Absolute error in this case are defined by the absolute difference between the average value and a specific value. The results of this test are summarized in Table.~\ref{tab2}. 

In test 2, a collection of CT scans of the chest-abdomen region was processed by a vertebrae segmentation algorithm. The segmentation results were used to automatically find the indices of distinct slices positioned exactly at the center point between T12-L1, L1-L2, L2-L3, L3-L4 and L4-L5 vertebrae centers. The scans were also processed by the proposed algorithm so that every slice of one of the five possible landmarks, was assigned with its estimated position value. Those values were then grouped by landmark and for each group, the average value, mean absolute error and median absolute errors were calculated. The results of this stage are summarized in Table.~\ref{tab3}.

\begin{table}[h]
\centering
\begin{tabular}{|c|c|c|c|c|}
\hline
Anatomical landmark & Mean position & MAE [cm]  & MdAE [cm]  & Number of scans\\
\hline

Superior aspect of the cortex & 0.4 & 0.2 [0.34] & 0.2 [0.34] & 28\\
Inferior cerebellum & 8.3 & 0.5 [0.85] & 0.4 [0.68] & 35\\
Thyroid & 15.3 & 0.7 [1.19] & 0.5 [0.85]  & 65\\
Apex of the lung & 15.6 & 0.6 [1.02] & 0.4 [0.68] & 71\\
Aortic arch & 19.2 & 0.7 [1.19] & 0.4 [0.68] & 72\\
Carina & 20.3 & 0.7 [1.19] & 0.4 [0.68] & 66\\
Pulmonary bifurcation & 20.8 & 0.8 [1.36] & 0.5 [0.85]  & 62\\
Inferior aspect of the heart & 27.1 & 1.1 [1.87] & 0.7 [1.19] & 55\\
Inferior aspect of the kidneys & 36.0 & 1.1 [1.87] & 0.9 [1.53] & 70\\
Superior ilium & 38.0 & 0.6 [1.02] & 0.5 [0.85]  & 70\\
Inferior ischium & 49.8 & 0.6 [1.02] & 0.4 [0.68] & 56\\
Patellas & 68.7 & 1.0 [1.7] & 0.9 [1.53] & 25\\

\hline
\end{tabular}
\vspace{5mm} 
\caption{Accuracy measurements of 12 visually detected anatomical landmarks. For each landmark we calculate the average position, Mean Absolute Error (MAE), and Median Absolute Error (MdAE) in normalized units and in cm. The rightmost column describes the number of scans included the landmark.}\label{tab2}
\end{table}

\begin{table}[h]
\centering
\begin{tabular}{|c|c|c|c|c|}
\hline
Anatomical landmark & Mean position & MAE [cm] & MdAE [cm] & Number of scans\\
\hline

T12-L1 & 30.8 & 0.7 [1.19] & 0.6 [1.02] & 1512\\
L1-L2 & 32.8 & 0.7 [1.19] & 0.5 [0.85]  & 1512\\
L2-L3 & 34.7 & 0.6 [1.02] & 0.5 [0.85]  & 1512\\
L3-L4 & 36.6 & 0.6 [1.02] & 0.5 [0.85]  & 1512\\
L4-L5 & 38.4 & 0.6 [1.02] & 0.5 [0.85]  & 1467\\

\hline
\end{tabular}
\vspace{5mm} 
\caption{Accuracy measurements of 5 automatically detected anatomical landmarks. For each landmark we calculate the average position, Mean Absolute Error (MAE), and Median Absolute Error (MdAE) in normalized units and in cm. The rightmost column describes the number of scans included the landmark.}\label{tab3}
\end{table}

\subsection{Comparison with other approaches}
The lack of a benchmark data-set for CT scan localization, limits the reliability of a comparative analysis of the different approaches. Nevertheless, we gathered the reported accuracy results of other existing approaches and summarised the error values in Table.~\ref{tab4}. Our MAE and MdAE values are the https://www.overleaf.com/project/651e8ac8f0ac68cded20369baverage of the 17 landmarks MAE and MdAE values of Table.~\ref{tab2} and~\ref{tab3}. Although our algorithm was tested on a variety of scan resolutions, protocols and regions with an emphasis on hard cases with strong artefacts, Table.~\ref{tab4} clearly shows our algorithm is more accurate the other existing approaches. 

In terms of processing time, the localization of an entire volume is ~0.4 seconds on a laptop with 16GB RAM, i7-7700HQ CPU, NVIDIA GTX-1050 GPU and windows 10 operating system. Average time for a CPU only mode (GPU is disabled) is ~0.73 seconds per scan. Since only 30 slices are actually processed - these results are invariant to the number of slices of the scan. For comparison, processing time was reported by \cite{ANATOMY-SPECIFIC} to be less than one minute per scan. \cite{CT_slice_localization} reported a range between 35 seconds to 140 seconds per scan with a processing time-accuracy trade-off. Non of the other referenced publications reported processing time.

\begin{table}[h]
\centering
\begin{tabular}{|c|c|c|c|c|}
\hline
Method & Year & MAE & MdAE\\
\hline
T. Emrich et al. \cite{CT_slice_localization} & 2010 & 2.83 & \\
A. Fernandez et al. \cite{Diffusion_methods} & 2014 & & 1.65\\
F. Graf et al. \cite{Position_prediction} & 2016 & 1.80 & \\
J. Guo et al. \cite{deep_learning_based} & 2017 & 1.69 & 1.04\\
\textbf{Ours} & 2023 & \textbf{1.18} & \textbf{0.88}\\

\hline
\end{tabular}
\vspace{5mm} 
\caption{Comparison with other approaches. Our approach is the most accurate in terms of MAE and MdAE.}\label{tab4}
\end{table}

\subsection{yield tests}
\begin{figure}[h]
\begin{center}
\includegraphics[width=0.6\textwidth]{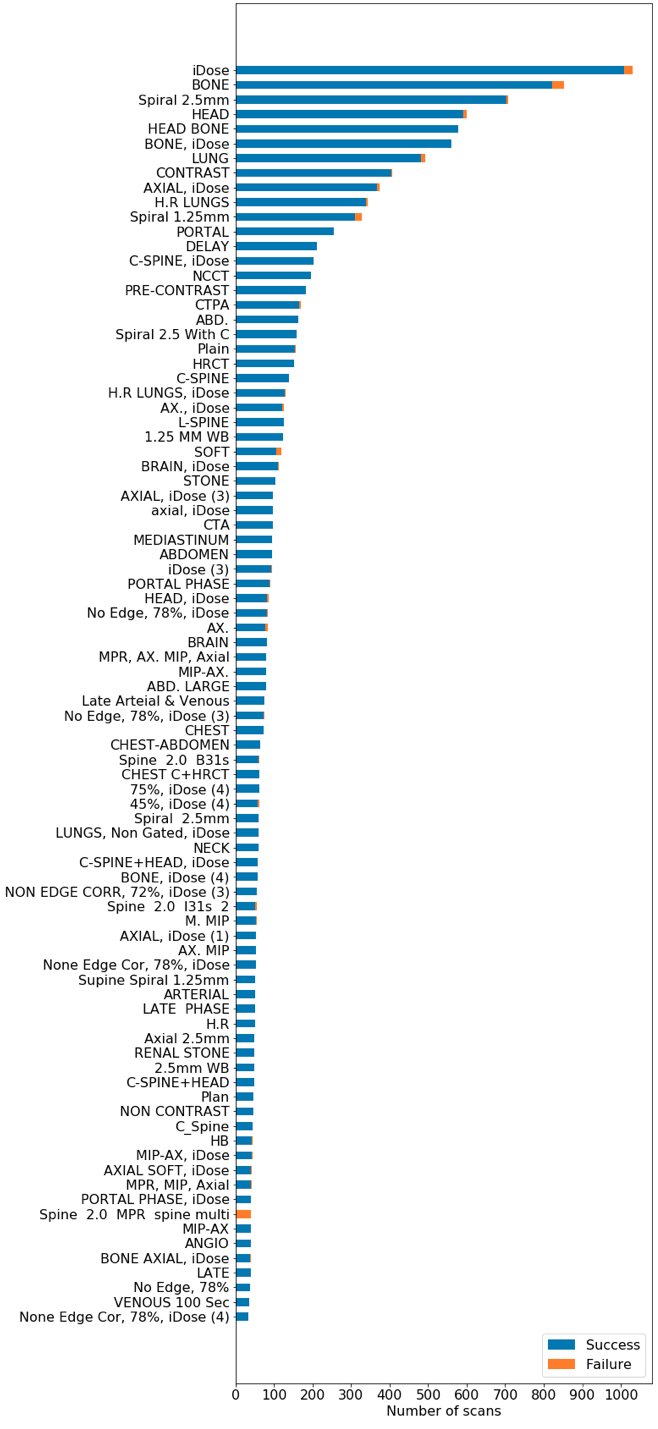}
\caption{Success/failure ratio for the 60 most frequent series descriptions} 
\label{yield}
\end{center}
\end{figure}

In order to set a rule based logic to determine whether a scan localization is reliable or should be disregarded, a collection of 4000 axial CT scans of various body parts, ages and protocols were randomly selected. Each scan of that collection was then localized by the above algorithm and a set of DICOM tags and fitting scores were assigned to each localization result. DICOM tags include 'pixel spacing' and 'spacing between slices'. The fitting score is the median absolute difference between the direct and the linear model based position of every slice out of the 30 processed slices. Each scan was also visualized as exemplified in Fig.~\ref{yield}. The author surveyed the entire collection and determined for each scan whether it was or wasn't reliably localized. The group of 4000 scans and their corresponding parameters was divided into two groups each composing 2000 scans. The first group was used to construct the rule based criteria which relies on the above properties to predict unreliable scans. This rule based criteria was then applied on the other group of scans in order to determine the algorithm's yield, and the rate of unreliable localization which weren't 'filtered out' by the set of rule. Out of the 2000 test scans, a yield of 97.5\% was reached with a single borderline scan localization which wasn't considered as unreliable by the exclusion criteria. In order to analyse the success rate of the algorithm on different scan protocols, the localization algorithm processed a larger collection of 21,500 unique scans of different body parts and protocols and the exclusion rules were applied on the localization results. Fig.~\ref{yield} shows the success/failure ratio for the 60 most frequent series descriptions found in our collection.

\begin{figure}[h]
\begin{center}
\includegraphics[width=0.9\textwidth]{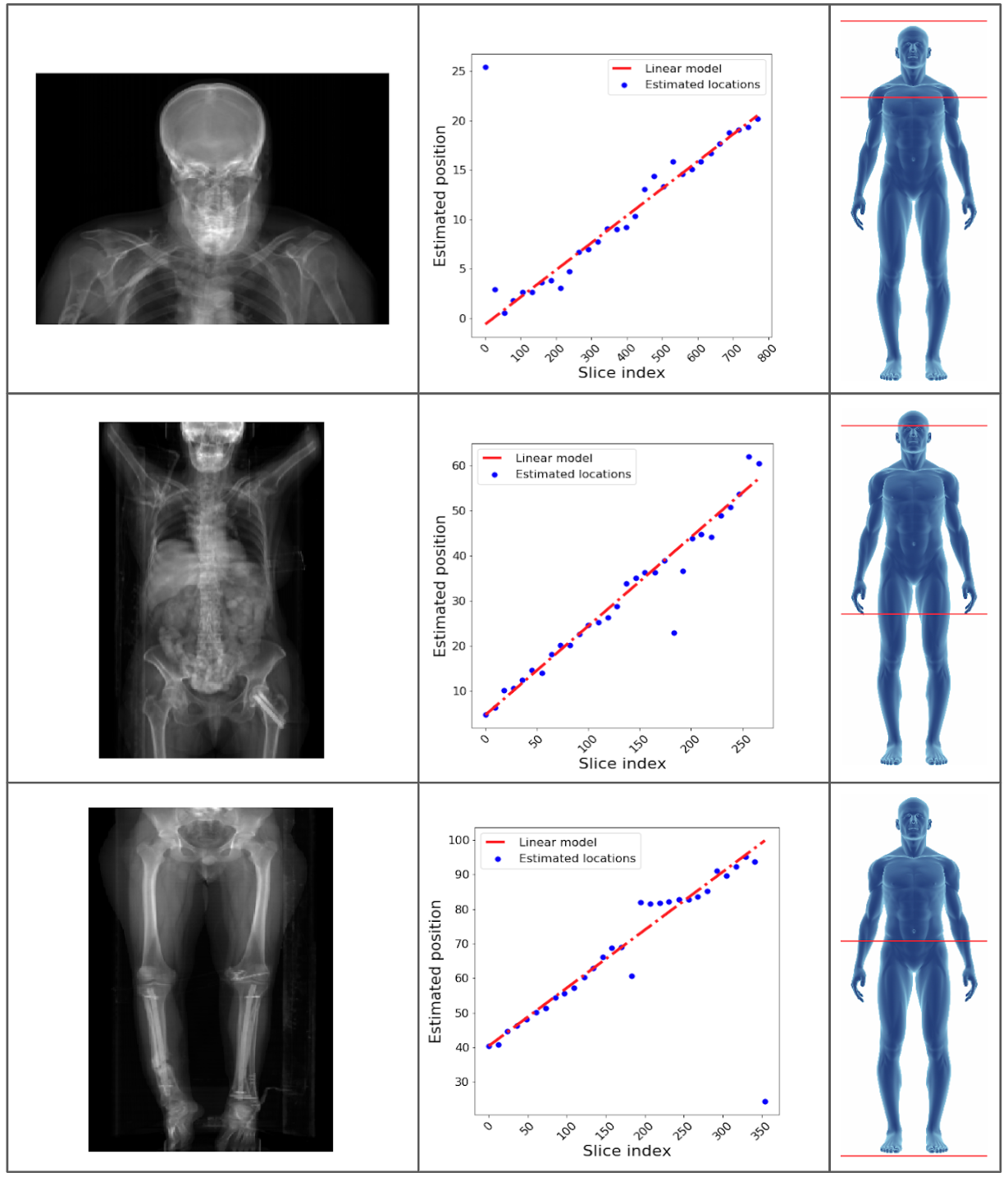}
\caption{Typical scan localization results. On the left - a 2D projection of the scan. On the center - positioning values versus slice index for the 30 processes slices (blue) and the estimated linear model based on those values (red). Notice that even large positioning errors do not affect the linear model and are replaced by it. On the right - a visualization of the scanned region based on the localization results.} 
\label{final_results}
\end{center}
\end{figure}

\section{Summary}

In this paper, a fast, accurate and robust method for axial CT scans was presented setting a new SOTA in terms of accuracy and running time. The linearity assumption of the slice index to normalized position seemed to be a valid assumption and had led to significant decrease in running time, an increase in accuracy and in robustness to strong artefacts. The principles presented in this work are not limited only to CT scans and can be applied to other imaging modalities as well where the most obvious candidate is MRI. Naturally, further research is required establish this supposition.

\section{Acknowledgement}
The author would like to thank Mr. Amir Bar and Dr. Ayelet Akselrod-Ballin for their support and to the entire 'Zebra medical Vision' team for constructive criticism of the proposed algorithm, the test analysis and the manuscript.

\bibliographystyle{unsrt}

\begin{thebibliography}{10}

\bibitem{health_at_glance-2017}
OECD.
\newblock {\em Health at a Glance}.
\newblock 2017.

\bibitem{DICOM-Quality}
Mark Oliver~Gueld et~al.
\newblock Quality of dicom header information for image categorization.
\newblock {\em Proc.SPIE}, 2002.

\bibitem{CT_slice_localization}
Tobias~Emrich et~al.
\newblock Ct slice localization via instance-based regression.
\newblock In {\em Medical Imaging: Image Processing}, 2010.

\bibitem{Comparing_axial_CT}
Feulner et~al.
\newblock Comparing axial ct slices in quantized n-dimensional surf descriptor space to estimate the visible body region.
\newblock {\em Computerized Medical Imaging and Graphics}, 2011.

\bibitem{Radial_Image_Descriptors}
Graf et~al.
\newblock 2d image registration in ct images using radial image descriptors.
\newblock 2011.

\bibitem{deep_learning_based}
Jiajia~Guo et~al.
\newblock A deep learning-based method for relative location prediction in {CT} scan images.
\newblock {\em CoRR}, 2017.

\bibitem{Position_prediction}
Graf et~al.
\newblock Position prediction in ct volume scans.
\newblock In {\em ICML, 2011}.

\bibitem{ANATOMY-SPECIFIC}
Holger~R. et~al.
\newblock Anatomy-specific classification of medical images using deep convolutional nets.
\newblock {\em CoRR}, 2015.

\bibitem{UNSUPERVISED}
Ke~Yan et~al.
\newblock Unsupervised body part regression using convolutional neural network with self-organization.
\newblock {\em CoRR}, 2017.

\bibitem{Estimating_the_body_portion}
Feulner et~al.
\newblock Estimating the body portion of ct volumes by matching histograms of visual words.
\newblock {\em SPIE}, 2009.

\bibitem{RANSAC}
Fischler et~al.
\newblock Random sample consensus: A paradigm for model fitting with applications to image analysis and automated cartography.
\newblock {\em Commun. ACM}, 1981.

\bibitem{Diffusion_methods}
Fernández et~al.
\newblock Diffusion methods for aligning medical datasets: Location prediction in ct scan images.
\newblock {\em Medical Image Analysis}, 2014.

\end{thebibliography}

\end{document}